\documentclass[doublecol]{epl2} 

\usepackage{amsmath}

\title{The Stationary Klein-Gordon Equation with a Delta-like Source: A Generalized Function Approach}
\shorttitle{The Stationary Klein-Gordon Equation with a Delta-like Source} 

\author{J.P. Ferreira\inst{1} \and F.E. Barone \and F.A. Barone\inst{1}}
\shortauthor{Ferreira \etal}

\institute{                    
  \inst{1} Instituto de F\'\i sica e Qu\'\i mica, Universidade Federal de Itajub\'a - Avenida BPS 1303, Itajub\'a, 37500-903, MG, Brazil
}

\abstract{This work aims to initiate a discussion on finding solutions to non-homoge\-neous differential equations in terms of generalized functions. For simplicity, we conduct the analysis within the specific context of the stationary Klein-Gordon equation with a point-like source, identifying a generalized function that solves such an equation and aligns with the solution obtained through the Fourier approach with dimensional regularization. In addition to being regular at the source singularity, a notable advantage of our solution is its presentation as a single expression, eliminating the need for piecewise definitions. The arguments presented here are applicable to a broader range of situations, offering a novel approach to addressing divergences in field theories using generalized functions. Moreover, we anticipate that the approach introduced in this work could provide a new method for handling Green functions regularized at coincident points, thereby simplifying the renorma\-lization process in a wide range of theories.}

\begin{document}

\maketitle

\section{Introduction}

Divergence-related problems in physics are widespread, especially in the realm of field theories. Notable examples in this context include point, linear, and surface-like field sources, along with their associated self-energies \cite{RMP1957,AnnPhys2002,KR,AnnPhys2007,JMP2008,BaroneHidalgo1,BaroneHidalgo2,BaroneHidalgoNoguieira,NPA2017,BaroneBaroneMedeiros,CS}. Beyond field sources, issues also arise when the fields themselves exhibit divergences, such as the electromagnetic fields of point-like sources (e.g., point-like charges, electric dipoles, magnetic dipoles) \cite{AJP2023,AJP1994,PR1946,PR1948,PRD1982A,JMP1989,PRD2019,PLA2013,AP2014}. Additionally, models with interacting fields may manifest divergences, necessita\-ting the redefinition of theory-inherent parameters through a process known as renormalization \cite{DasTQC,Itzykson,Barcelos,GreinerQED,Kaku,Weinberg}.

Some of the previously mentioned divergences are integrable and can be expressed in terms of distributions, specifically Dirac delta functions and Heaviside step functions. This allows us to systematically address these divergences. Examples include the charge density associated with point-like, linear, and surface distributions, as well as fields associated with certain point-like sources, such as the electric dipole field and the magnetic dipole field \cite{DasEletro,Zangwill,Jackson,Frenkel}. Notably, there are instances of divergent fields produced by linear source distributions, such as the magnetic field of Dirac strings \cite{Felsager,BaroneHelayel,CamiloBaroneHelayel}. Indeed, the use of distribution theory provides an effective means of dealing with idealized physical objects and managing singularities that underlie many problems in physics.

Here we highlight the direct employment of distributions for the control of divergences in quantum field theories \cite{qftdist}. We also point out that there are many ways to represent a Dirac delta function or a Heaviside step function in terms of generalized functions, in special we mention the ones obtained from series of functions \cite{BJP2021,BJPamaku}.

As far as the authors are aware, the theory of distributions was initially discussed in the 19th century by Poisson, Fourier, and Cauchy in the context of solving partial differential equations and functional analysis \cite{historydist}. One of the earliest formal formulations of $\delta$-functions appeared in the work of Heaviside, who, in the context of telegraphy problems, argued that $\delta$-functions should be the derivatives of the step unity function \cite{heavisidehist}. Independently, Dirac developed his $\delta$-function formalism in the realm of quantum mechanics and functional analysis to provide an interpretation of the unitary element in functional space, drawing an analogy with the Kronecker delta \cite{deltazerotoinf}. The formal mathematical construction of distributions was later introduced by Schwartz \cite{laurent,SMA15}, where distributions are presented in an abstract sense within the context of functional analysis.

With Schwartz's theory, the formal use of distributions in field theories became possible for describing localized sources and external potentials. However, challenges arose when dealing with non-linear quantities involving distributions, such as $\delta^2$ or $\Theta \delta$ (where $\Theta$ is the step unity function), which were ill-defined within the framework. This issue was addressed by the formalism proposed by Coulombeau \cite{colombeauLivro,colombeau,egorov,vickers,steinbauer,dannon}, who extended Schwartz's theory. The advancements in Coulombeau's approach enabled the definition of functions of distributions \cite{egorov,dannon}.

With the advancements in the theory of distributions, a natural question arises: Can a distribution serve as a solution to a differential equation? In typical physics problems, it is common to encounter delta-like field sources or delta-like potentials coupled to a field. However, it is less conventional to have a distribution as a field solution. In this paper, we engage in a discussion on this topic in connection with the stationary Klein-Gordon equation with a point-like source. The considered source assumes a role analogous to that of an electric charge in electromagnetism.

In this paper, we demonstrate the existence of a solution comprised of generalized functions for the stationary Klein-Gordon equation with a delta-like source centered at the origin. The solution reproduces the Yukawa potential outside the origin, and, moreover, at the origin, aligns with the one obtained through dimensional regularization of the solution in Fourier integral form. It is worth mentioning that the differential equation considered in this paper is related to the modified Helmholtz differential operator in three dimensions.

We chose to address this specific problem for several reasons: the calculations are not overly complicated; it holds high relevance in numerous physics problems, and, most importantly, the arguments can be extended to the Green’s function for the Klein-Gordon field as well as other fields, especially the Maxwell field. This last issue is of special interest since the evaluation of Green functions at coincident points plays a central role in the renormalization scheme of interacting field theories \cite{Kaku,Weinberg}. 

Additionally, the approach proposed in this work can run parallel to the standard dimensional regularization scheme, but offers the advantage of explicitly handling the subtraction of divergent contributions. Moreover, the proposed method leads to a single expression without the need for piecewise definitions.

In the first part of the paper, we offer a concise initial review and discussion on the regularization of the Yukawa potential as a solution to the stationary Klein-Gordon equation with a delta-like source. Subsequently, we proceed with the Fourier analysis of the stationary Klein-Gordon equation under the influence of a delta-like source. Utilizing dimensional regularization, we determine the value of its associated solution at the delta singularity. After this review discussion, we propose a new solution for the stationary Klein-Gordon equation with a delta-like source in terms of distributions and analyze its properties. This solution is square integrable, regular at the source singularity, and equal to the Yukawa potential outside the source singularity. Furthermore, it aligns with the solution obtained from the Fourier integral with dimensional regularization. We end this letter with some conclusions and final remarks, offering a discussion regarding the use of a similar approach to other problems in physics, mainly those involving Green functions.

\section{The Yukawa potential and the Fourier approach with dimensional regularization}
\label{secaoeqdiferencial}

In this section, we provide a brief review of the stationary Klein-Gordon equation with a delta-like source. The results are not novel, and we include them for the sake of completeness, ensuring the paper is self-contained.

Let us start by considering the homogeneous stationary Klein-Gordon equation with radial symmetry,
\begin{equation}
\label{KGlivre}
\Big(\nabla^2 - m^2 \Big)\phi(r)=\frac{1}{r}\frac{d^{2}}{dr^{2}}[r\phi(r)]-m^{2}\phi(r)=0\ .
\end{equation}
where $r$ is the radial coordinate in standard spherical coordinates.

A first attempt to find solutions for (\ref{KGlivre}) leads to
\begin{eqnarray}
\label{15}
\phi_{-}(r)=\frac{e^{-m r}}{r}\ \ ,\ \ 
\phi_{+}(r)=\frac{e^{+m r}}{r}\ .
\end{eqnarray}

The second solution (\ref{15}) is not square integrable and diverges at the infinity, $r\to\infty$. The first equation (\ref{15}) is the so called Yukawa potential, which is square-integrable and finite as $r\to\infty$. From now on, we shall discard the second solution (\ref{15}) and consider only the first one in our analysis.

In fact, the first equation (\ref{15}) is a solution for (\ref{KGlivre}) just outside the origin, where $r\neq0$. However, at the origin this solution tends to infinity. To analyze its behavior at this point, it is necessary to tame this divergence. To achieve this, we employ a regularization scheme by introducing a regularization parameter $\epsilon$, as follows \cite{Felsager}
\begin{equation}
\label{Yukawaregularizado}
\frac{e^{-m r}}{r}\to \frac{e^{-m r}}{\sqrt{r^{2}+\epsilon^{2}}}
\end{equation}
where is implicit that, at the conclusion of any calculation, the limit $\epsilon\to0$ must be taken. In fact, in this limit, the regularized solution (\ref{Yukawaregularizado}) precisely becomes the Yukawa potential given in the first equation (\ref{15}).

Substituting (\ref{Yukawaregularizado}) in the right hand side of (\ref{KGlivre}) we have
\begin{eqnarray}
\label{solucaonaequacao}
\frac{1}{r}\frac{d^{2}}{dr^{2}}\Bigg[r\frac{e^{-m r}}{\sqrt{r^{2}+\epsilon^{2}}}\Bigg]-m^{2}\frac{e^{-m r}}{\sqrt{r^{2}+\epsilon^{2}}}=\cr\cr
=\Bigg[-\frac{3\epsilon^{2}}{(r^{2}+\epsilon^{2})^{5/2}}-2m\frac{\epsilon^{2}}{r(r^{2}+\epsilon^{2})^{3/2}}\Bigg]e^{-m r}\ .
\end{eqnarray}

The exponential term in the second line of (\ref{solucaonaequacao}) is square integrable and finite across the entire domain $\Re^{3}$. The first term inside the brackets on the right-hand side represents a delta function \cite{Felsager}. It is important to note that the limit $\epsilon\to0$ manifests distinct behaviors when taken at points outside the origin and at $r=0$, specifically
\begin{eqnarray}
\label{limitesprimeirotermo}
    \lim_{\epsilon=0}\Bigg(-\frac{3\epsilon^{2}}{(r^{2}+\epsilon^{2})^{5/2}}\big|_{r=0}\Bigg) &=& \lim_{\epsilon=0}\Bigg(-\frac{3}{\epsilon^3} \Bigg)=\infty\\ \nonumber
    \lim_{\epsilon=0}\Bigg(-\frac{3\epsilon^{2}}{(r^{2}+\epsilon^{2})^{5/2}}\big|_{r\not=0}\Bigg) &=& 0.
\end{eqnarray}

The properties (\ref{limitesprimeirotermo}) are the ones expected for a delta function. Moreover, it can be shown that the divergence exhibited by this term in the limit (\ref{limitesprimeirotermo}) at the origin is integrable \cite{Felsager}. This can be demonstrated by integrating the first term inside the brackets of the last line of (\ref{solucaonaequacao}) across the entire space and then taking the limit $\epsilon\to0$, as follows
\begin{eqnarray}
\label{tgb1}
\int d^{3}{\bf r}\frac{-3\epsilon^{2}}{(r^{2}+\epsilon^{2})^{5/2}}=4\pi\int_{0}^{\infty}dr\frac{-3r^{2}\epsilon^{2}}{(r^{2}+\epsilon^{2})^{5/2}}=-4\pi\ .
\end{eqnarray}

The integral (\ref{tgb1}) does not depend on $\epsilon$, so its limit as $\epsilon\to0$ is also equal to $-4\pi$. Therefore, this demonstrates that the divergence of the first term inside the brackets in the second line of (\ref{solucaonaequacao}) in the limit $\epsilon\to0$ is integrable. Consequently, this term is, indeed, proportional to a Dirac delta function, namely
\begin{equation}
-\frac{3\epsilon^{2}}{(r^{2}+\epsilon^{2})^{5/2}}\to-4\pi\delta^{3}({\bf r}) \ .
\end{equation}

The limit $\epsilon\to0$ of the last term inside brackets in (\ref{solucaonaequacao}) is ill-defined at the origin and vanishes for points where $r\neq0$,
\begin{eqnarray}
\label{limitessegundotermo}
    \lim_{\epsilon=0}\Bigg(\frac{\epsilon^{2}}{r(r^{2}+\epsilon^{2})^{3/2}}\big|_{r=0}\Bigg) &=& \frac{1}{0}\lim_{\epsilon=0}\Bigg(\frac{1}{\epsilon}\Bigg)=\infty\\ \nonumber
    \lim_{\epsilon=0}\Bigg(\frac{\epsilon^{2}}{r(r^{2}+\epsilon^{2})^{3/2}}\big|_{r\not=0}\Bigg) &=& 0.
\end{eqnarray}
Furthermore, its integral across the entire space vanishes in the limit $\epsilon\to0$ since
\begin{equation}
\int d^{3}{\bf r}\frac{\epsilon^{2}}{r(r^{2}+\epsilon^{2})^{3/2}}=\epsilon\ .
\end{equation}
Due to the fact that this term is always positive across the entire space, we can conclude that it equals zero in the limit $\epsilon\to0$.

Collecting the previous results, we can rewrite 
\begin{eqnarray}
\label{finalequacaoYukawa}
\Big(\nabla^2 - m^2 \Big)\frac{e^{-m r}}{r}\to \lim_{\epsilon=0}\Big(\nabla^2 - m^2 \Big)\frac{e^{-m r}}{\sqrt{r^{2}+\epsilon^{2}}}=\cr\cr
=-4\pi\delta^{3}({\bf r})e^{-m r}=-4\pi\delta^{3}({\bf r})
\end{eqnarray}
where, in the last step, we used the property $\delta^{3}(\mathbf{r})f(\mathbf{r})=\delta^{3}(\mathbf{r})f(\mathbf{r}=0)$, which is valid for any square integrable and finite function across the entire $\Re^{3}$.

The result (\ref{finalequacaoYukawa}) shows that the Yukawa potential, as given in the first equation (\ref{15}), is not the solution for the homogeneous stationary Klein-Gordon equation (\ref{KGlivre}) across the entire space. However, it is a solution for this equation with a stationary delta-like source concentrated at the origin, as expected.

Now let us present a brief review of the Fourier approach for the stationary Klein-Gordon equation with a delta-like source,  
\begin{equation}
\label{14}
\Big(\nabla^2 - m^2 \Big)\phi(\mathbf{r})=-4\pi \delta^3(\mathbf{r})\ .
\end{equation}

We will conduct an analysis using dimensional regularization to achieve a finite value for the field solution at the singular source.

Applying a Fourier approach to solve equation (\ref{14}) leads to the integral
\begin{equation}
\label{integralFourier}
\phi_{F}(r)=\frac{1}{2\pi^{2}}\int d^{3}{\bf k}\frac{\exp(-i{\bf k}\cdot{\bf r})}{{\bf k}^{2}+m^{2}}\ .
\end{equation}

This fact can be verified by substituting (\ref{integralFourier}) into the left-hand side of (\ref{14}) and employing the Fourier representation of the Dirac delta function where
\begin{equation}
\delta^{3}({\bf r})=\frac{1}{(2\pi)^{3}}\int d^{3}{\bf k}\exp(-i{\bf k}\cdot{\bf r})\ .
\end{equation}

The integral in (\ref{integralFourier}) can be solved by employing spherical coordinates \cite{Kaku}. For points where ${\bf{r}}\neq0$, we have,
\begin{equation}
\phi_{F}(r)=\frac{1}{2\pi^{2}}\int d^{3}{\bf k}\frac{\exp(-i{\bf k}\cdot{\bf r})}{{\bf k}^{2}+m^{2}}=\frac{e^{-mr}}{r}=\phi_{-}(r)\ ,\ {\bf{r}}\not=0
\end{equation}
where, in the last step, we used the definition (\ref{15}).

The integral (\ref{integralFourier}) is evidently divergent at the origin, where ${\bf r}=0$. However, we can establish a well-defined value for it through analytic continuation. For this purpose, we use the fact that in $D$ dimensions \cite{Kaku},
\begin{eqnarray}
\label{integralextanalitica}
\phi_{F}(r=0)\to\frac{1}{2\pi^{2}}\lim_{D=3}\left(\int d^{D}{\bf k}\frac{1}{{\bf k}^{2}+m^{2}}\right)\cr\cr
=\frac{1}{2\pi^{2}}\lim_{D=3}\left(\frac{\pi^{D/2}\Gamma(1-D/2)}{(m^{2})^{1-D/2}}\right)\ ,
\end{eqnarray}
so the solution (\ref{integralFourier}) at the origin can be obtained by taking $D=3$ in (\ref{integralextanalitica}),
\begin{equation}
\phi_{F}(r=0)=-m\ .
\end{equation}

Collecting terms, the Fourier approach with the analytic continuation for the origin leads to the solution
\begin{eqnarray}
\label{49}
\phi_{F}(r)=
    \begin{cases}
    \phi_{-}(r)=\frac{e^{-mr}}{r},\; r\neq0,\\
    -m,\; r=0\ .
    \end{cases}
\end{eqnarray}

At this point, some comments are in order. The Fourier approach used to solve equation (\ref{14}) leads to the integral (\ref{integralFourier}), which yields the Yukawa potential. However, this potential is not defined at the origin (due to the source singularity). The dimensional regularization employed in result (\ref{integralextanalitica}) is based on the analytical extension of the underlying integral, where the original (divergent) function is replaced by its analytical extension. In this case, the analytical extension is well-defined at the source singularity. Therefore, establishing result (\ref{49}) requires not only the Fourier approach but also a regularization scheme (the dimensional one). Furthermore, the Fourier approach deals only with square-integrable functions, and thus does not encompass the positive exponential solution in the second equation (\ref{15}).

\section{A solution as a generalized function}
\label{solucaoDistribuicao}

Establishing a solution to the stationary Klein-Gordon equation with a delta-like source concentrated at the origin, which remains finite there, is crucial for addressing various issues in field theories. However, the use of analytic extension may be considered somewhat artificial because every regularization scheme involves subtracting divergent quantities, and dimensional regularization, despite its sophistication, obscures this subtraction by employing analytic extension. The arguments leading to result (\ref{49}) rely on dimensional regularization of a divergent integral in (\ref{integralFourier}), thus keeping the aforementioned subtraction implicit in this case.

It would be beneficial to propose a regularization approach that runs parallel to dimensional regularization but explicitly handles the subtraction of divergent contributions. Furthermore, having a solution that is well-behaved at the origin and represented by a single expression (thereby avoiding piecewise definitions like in (\ref{49})) would be highly advantageous. Achieving such a solution is not feasible with single functions alone but can be accomplished by employing generalized functions or distributions, as demonstrated in this section.

The fundamental idea behind generalized functions is to obtain a function that exhibits certain desired properties (which, in some cases, cannot be satisfied by any ordinary function) only under a specific limit.

We begin by defining the function
\begin{equation}
\label{deff}
f(x,\epsilon):=\frac{1}{2}\Bigg[1+\mbox{Erf}\Bigg(\frac{x}{\sqrt{2}\epsilon}\Bigg)\Bigg]\ ,
\end{equation}
where $\mbox{Erf}(x)$ stands for the Gauss error function. 

It can be shown that $f(x,\epsilon)$ exhibits the following property
\begin{equation}
\label{propriedades}
    \lim_{\epsilon=0^{+}}f(x,\epsilon)
		=\begin{cases}
    1,\; x>0,\\
    \frac{1}{2},\;x=0,\\
    0,\; x<0\ .
		\end{cases}
\end{equation}
So, $f(x,\epsilon)$ is a representation for the Heaviside function in terms of a generalized function in the sense that
\begin{equation} 
\label{representacaoTheta}
\Theta(x)=\lim_{\epsilon=0^{+}}f(x,\epsilon)\ .
\end{equation}

Here we point out that in this work we define the Heaviside in such a way that $\Theta(0)=1/2$. 

The derivative of the Heaviside function is the Dirac delta function, namely
\begin{equation}
\frac{d\Theta(x)}{dx}=\delta(x)\ .
\end{equation}
Taking the derivative of expression (\ref{deff}) we obtain the well known gaussian representation for the Dirac delta function,
\begin{equation}
\frac{df(x,\epsilon)}{dx}=\frac{1}{\sqrt{2\pi}\epsilon}\exp\Bigg(-\frac{x^{2}}{2\epsilon^{2}}\Bigg)\ .
\end{equation}

Now, we make the {\it ad-hoc} assertion that the generalized function
\begin{eqnarray}
\label{50}
         \Phi(r,\epsilon)&=&f(r,\epsilon)\phi_{-}(r)- f(-r,\epsilon)\phi_{+}(r)\cr\cr
				 &=&\frac{1}{2}\Bigg[1+\mbox{Erf}\Bigg(\frac{r}{\sqrt{2}\epsilon}\Bigg)\Bigg]\frac{e^{-m r}}{r}\cr\cr
				&\ &-\frac{1}{2}\Bigg[1+\mbox{Erf}\Bigg(\frac{-r}{\sqrt{2}\epsilon}\Bigg)\Bigg]\frac{e^{m r}}{r}\ ,
\end{eqnarray}
where we have employed the definitions (\ref{15}) and (\ref{deff}), is indeed a solution to the equation (\ref{14}). Furthermore, it is equivalent to (\ref{49}) in the sense that
\begin{eqnarray}
\label{afirmacoes}
\lim_{\epsilon=0^{+}}\Big(\nabla^2 - m^2 \Big)\Phi(r,\epsilon)=-4\pi\delta^3(\bf{r})\ ,\cr\cr
\lim_{\epsilon=0^{+}}\Phi(r,\epsilon)=\phi_{F}(r)
\end{eqnarray}
with $\phi_{F}(r)$ defined in (\ref{49}).

In fact, for the definition (\ref{deff}), we must choose a representation for the Heaviside function that tames the divergence of the function $\phi_{+}(r)=e^{mr}/r$ as $r$ approaches infinity ($r\to\infty$). The function $f(r,\epsilon)$ is suitable for this purpose. One might wonder why Heaviside functions $\Theta(r)$ and $\Theta(-r)$ were not used instead of $f(r,\epsilon)$ and $f(-r,\epsilon)$ in definition (\ref{50}). The reason lies in the fact that if Heaviside functions had been used, any representation for them in terms of generalized functions could be employed, not just the one defined in (\ref{deff}). This wouldn't pose a problem for the first term on the right-hand side of (\ref{deff}), which contains the exponential $e^{-m r},$ but it could be problematic for the second term due to the exponential $e^{m r},$ as mentioned.

Besides, expression (\ref{50}) is a combination of the functions (\ref{15}), which are solutions for the Klein-Gordon equation out of the origin, one with the negative argument exponential (the Yukawa potential) and the other with the positive argument exponential. These two functions diverge at the origin with the same behavior, and the definition (\ref{50}) is constructed in such a way to make divergences of both functions (\ref{15}) to cancel each other. All features of definitions (\ref{50}), (\ref{deff}) and (\ref{15}) make the assertions (\ref{afirmacoes}) to be true, as will be shown.

It is straightforward to prove the second assertion (\ref{afirmacoes}). For $r>0$ the distribution $(\ref{50})$ reduces to
\begin{equation}
\label{phiforaorigem}
\lim_{\epsilon\to0^{+}}\Phi(r>0,\epsilon)=\phi_{-}(r>0)=\frac{e^{-m r}}{r}\ ,\ r>0\ ,
\end{equation}
where we used expression (\ref{deff}) and the properties (\ref{propriedades}).

To evaluate eq. (\ref{50}) for $r=0$ we must use definition (\ref{deff}), what leads to
\begin{eqnarray}
\label{phiorigem}
\lim_{\epsilon\to0^{+}}\Phi(r>0,\epsilon)=\frac{1}{2}\Biggl(\frac{e^{-m r}}{r}-\frac{e^{m r}}{r}\Biggr)\Big|_{r=0}\cr\cr
=\frac{1}{2}[-2m+{\cal O}(r^2)]|_{r=0}=-m\ .
\end{eqnarray}

From results (\ref{phiforaorigem}) and (\ref{phiorigem}), we can see that (\ref{50}) is indeed a representation of the result (\ref{49}) in terms of generalized functions, which proves the second assertion (\ref{afirmacoes}).

Now, we must prove the first statement (\ref{afirmacoes}) by showing that (\ref{50}) is really a solution for (\ref{14}).

Notice that the function (\ref{50}) does not diverge at the origin. Therefore, there is no need to regularize the denominator $1/r$. Substituting (\ref{50}) into the left-hand side of (\ref{14}) and performing some simple manipulations, we have
\begin{eqnarray}
\label{qaz1}
\Big(\nabla^2 - m^2 \Big)\Phi(r,\epsilon)=
\frac{1}{r}\frac{d^{2}}{dr^{2}}\Big[r\Phi(r,\epsilon)\Big]-m^{2}\Phi(r,\epsilon)\cr\cr
=-\frac{2m}{\sqrt{2\pi}\epsilon}e^{-r^{2}/(2\epsilon^{2})}\frac{\big(e^{-mr}-e^{mr}\big)}{r}\cr\cr
+\frac{\big(e^{-mr}+e^{mr}\big)}{r}\frac{d}{dr}\Bigg(\frac{1}{\sqrt{2\pi}\epsilon}e^{-r^{2}/(2\epsilon^{2})}\Bigg)\ .
\end{eqnarray}

Let us focus on the term in the second line of equation (\ref{qaz1}). It is given by a Gaussian representation of the Dirac delta function in the argument $r$ multiplied by a function of $r$. This Gaussian representation controls the divergence exhibited by the exponential with a positive argument. For points outside the origin, with $r \neq 0$, the limit $\epsilon \to 0$ of this term goes to zero. For the origin, the term in parentheses equals to $-2mr$, and the Gaussian diverges in the limit $\epsilon \to 0$ due to the pre-exponential factor $1/\epsilon$. So, in summary,
\begin{eqnarray}
\lim_{\epsilon=0}\Bigg[\frac{-2m}{\sqrt{2\pi}\epsilon}e^{-r^{2}/(2\epsilon^{2})}\frac{\big(e^{-mr}-e^{mr}\big)}{r}\Bigg]\Bigg|_{r\not=0}=0\ ,\cr\cr
\lim_{\epsilon=0}\Bigg[-\frac{2m}{\sqrt{2\pi}\epsilon}e^{-r^{2}/(2\epsilon^{2})}\frac{\big(e^{-mr}-e^{mr}\big)}{r}\Bigg]\Bigg|_{r=0}=\cr\cr
=\frac{4m^{2}}{\sqrt{2\pi}}\lim_{\epsilon=0}\frac{1}{\epsilon}=\infty\ .
\end{eqnarray}

Therefore, in the limit $\epsilon=0$, the second line of equation (\ref{qaz1}) vanishes in the entire $\Re^3$ except at the origin, where it diverges. We have to investigate whether this divergence at this single point characterizes this term as a Dirac delta function or not. If this divergence is integrable and not equal to zero, this term can be proportional to a representation of the Dirac delta function. If this divergence is integrable and equal to zero, this term can be taken as equal to zero because this single-point divergence shall not bring any contribution in any circumstance. Additionally, since this term is non-negative throughout $\Re^3$, its integral is equal to zero if, and only if, it vanishes. This integration can be done straightforwardly,
\begin{eqnarray}
\label{qaz2}
-\frac{2m}{\sqrt{2\pi}\epsilon}\int d^{3}{\bf r}\ e^{-r^{2}/(2\epsilon^{2})}\frac{\big(e^{-mr}-e^{mr}\big)}{r}\cr\cr
=\int_{0}^{\infty}dr\ \frac{2^{3}\pi mr}{\sqrt{2\pi}\epsilon}e^{-r^{2}/(2\epsilon^{2})}\big(e^{mr}-e^{-mr}\big)\cr\cr
=2^3\pi m^2\sqrt{\epsilon}\ e^{\epsilon m^{2}/2}\ .
\end{eqnarray}
In the limit $\epsilon=0$, the right-hand side of (\ref{qaz2}) equals zero, so the integrand is also equal to zero.

Now, we turn our attention to the third line of equation (\ref{qaz1}). In this case, we have the derivative of the Dirac delta function (in the Gaussian representation) multiplied by a function of $r$. The Gaussian controls the divergence of the exponential function at infinity. For the limit $\epsilon=0$, we must consider the points outside the origin and the origin separately,
\begin{eqnarray}
\label{qaz3}
\lim_{\epsilon=0}\Bigg[\frac{d}{dr}\Bigg(\frac{1}{\sqrt{2\pi}\epsilon}e^{-r^{2}/(2\epsilon^{2})}\Bigg)\frac{\big(e^{-mr}+e^{mr}\big)}{r}\Bigg]\Bigg|_{r\not=0}\cr\cr
=-\lim_{\epsilon=0}\Bigg[\frac{1}{\sqrt{2\pi}\epsilon^{3}}e^{-r^{2}/(2\epsilon^{2})}\big(e^{-mr}+e^{mr}\big)\Bigg]\Bigg|_{r\not=0}=0\ ,\cr\cr\cr
\lim_{\epsilon=0}\Bigg[\frac{d}{dr}\Bigg[\frac{1}{\sqrt{2\pi}\epsilon}e^{-r^{2}/(2\epsilon^{2})}\Bigg]\frac{\big(e^{-mr}+e^{mr}\big)}{r}\Bigg]\Bigg|_{r=0}\cr\cr
=-\lim_{\epsilon=0}\Bigg(\frac{2}{\sqrt{2\pi}\epsilon^{3}}\Bigg)\Bigg|_{r=0}=\infty\ .
\end{eqnarray}

The result (\ref{qaz3}) indicates that the limit $\epsilon=0$ of the third line in (\ref{qaz1}) equals to zero outside the origin and diverges at the origin. The integral of this term is given by
\begin{eqnarray}
\int d^{3}{\bf r}\frac{d}{dr}\Bigg(\frac{1}{\sqrt{2\pi}\epsilon}e^{-r^{2}/(2\epsilon^{2})}\Bigg)\frac{\big(e^{-mr}+e^{mr}\big)}{r}\cr\cr
=-\int_{0}^{\infty}dr\ \frac{4\pi r^2}{\sqrt{2\pi}\epsilon^{3}}e^{-r^{2}/(2\epsilon^{2})}\big(e^{-mr}+e^{mr}\big)\cr\cr
=-4\ e^{\epsilon^{2}m^{2}/2}(1+\epsilon^{2}m^{2})\pi\ .
\end{eqnarray}
In the limit $\epsilon=0$ the above integral equals to $-4\pi$.

Therefore, the third line of equation (\ref{qaz1}), in the limit $\epsilon=0$, is equal to zero outside the origin and diverges at the origin. This divergence is not only integrable, but, more importantly, its integral over the entire $\Re^3$ is non-vanishing as well. In fact, this term is a representation of the Dirac delta function multiplied by the factor $-4\pi$.

Collecting the previous results, we can see that the right hand side of (\ref{qaz1}) is given by $-4\pi\delta^{3}({\bf r})$, therefore,
\begin{equation}
\Big(\nabla^2 - m^2 \Big)\Phi(r,\epsilon)=-4\pi\delta^{3}({\bf r})
\end{equation}
where it is implicit that the limit $\epsilon=0$, what proves that the generalized function (\ref{50}) is indeed a solution for the stationary Klein-Gordon equation with a point-like delta function.

In the definition (\ref{50}), we have used a special representation for the Heaviside step function in such a way as to tame the divergence introduced by the exponential with a positive argument. Not all representations of the Heaviside step function would have this property. That is why we avoided using the Heaviside step function in a general sense in our calculations and opted for a specific representation instead.

We hope that the main ideas exposed in this work could be a contribution to the use of distributions in a broader sense, particularly in the context of solutions to significant differential equations in Physics in connection with standard regularization techniques. Perhaps, such an approach could be employed to treat Green functions as well as $n$-point functions in field theories. We leave this topic as an open question to be explored in a broader context.

\section{Conclusions and final remarks}
\label{conclusoes}

We have revisited the stationary Klein-Gordon equation with a delta-like source concentrated at the origin. We have shown that there is a solution for this equation in the context of generalized functions (distributions) that recovers the Yukawa potential outside the origin and is finite at the origin. The proposed solution yields the same result as the one obtained from the Fourier integral, employing dimensional regularization at the origin, but is defined as a single function over the entire $\Re^{3}$, without the need to resort to piecewise expressions.

We hope that the discussion developed in this work could provide a systematic way to establish field solutions in terms of generalized functions in larger domains than the ones obtained when working solely with ordinary functions. This approach could be particularly relevant for dealing with physical quantities that require renormalization, especially in issues related to field theories.

Furthermore, we hope that the approach presented in this paper can be extended to more generalized situations, such as non-stationary differential equations and differential equations for other types of fields. Moreover, maybe the methods proposed in this paper might also be applicable to the development of an alternative regularization program based on generalized functions. We leave these points as open questions to be addressed by interested readers in future papers.

\acknowledgments F.A. Barone thanks to CNPq under the grant 313426/2021-0. J.P. Ferreira thanks to CNPq for invaluable financial support.


\begin{thebibliography}{0}

\bibitem{RMP1957} S. Deser, Rev. Mod. Phys. {\bf 29}, 417 (1957).

\bibitem{AnnPhys2002} M. Marino, Ann. Phys. {\bf 301}, 85 (2002).

\bibitem{KR} F.A. Barone, L.M. De Moraes, J.A. Helay\"el-Neto, Phys. Rev. D {\bf 72}, 105012 (2005). (Erratum: Phys. Rev. D {\bf 73}, 089901 (2006)).

\bibitem{AnnPhys2007} K. Lechner and P.A. Marchetti, Ann. Phys. {\bf 322}, 1162 (2007).

\bibitem{JMP2008} A. Gsponer, J. Math. Phys. {\bf 49}, 102901 (2008).

\bibitem{BaroneHidalgo1} F.A. Barone, G. Flores-Hidalgo, Phys. Rev. D {\bf 78}, 125003 (2008).

\bibitem{BaroneHidalgo2} F.A. Barone and G. Flores-Hidalgo, Braz. J. Phys. {\bf 40}, 188 (2010).

\bibitem{BaroneHidalgoNoguieira} F.A. Barone, G. Flores-Hidalgo and A. A. Nogueira, Phys. Rev. D {\bf 91}, 027701 (2015).

\bibitem{NPA2017} R. Holliday, R. McCarty, B. Peroutka and K. Tuchin, Nucl. Phys. A {\bf 957}, 406 (2017).

\bibitem{BaroneBaroneMedeiros} M.F.X.P. Medeiros, F.E. Barone, F.A. Barone, Eur. Phys. J. C {\bf 78}, 12 (2018).

\bibitem{CS} L.H.C. Borges, F.E. Barone, C.C.H. Ribeiro, H.L. Oliveira, R.L. Fernandez, F.A. Barone, Eur. Phys. J. C {\bf 80}, 238 (2020).






\bibitem{AJP2023} T. Boyer, Am. J. Phys. {\bf 91}, 74 (2023).

\bibitem{AJP1994} O.D. Jefimenko, Am. J. Phys. {\bf 62}, 79 (1994)

\bibitem{PR1946} M. Sch\"onberg, Phys. Rev. {\bf 69}, 211 (1946).

\bibitem{PR1948} P. Havas, Phys. Rev. {\bf 74}, 456 (1948).

\bibitem{PRD1982A} R.E. Kates, Phys. Rev. D {\bf 25}, 2487 (1982).

\bibitem{JMP1989}  A. Lozada, J. Math. Physics {\bf 30}, 1713 (1989).

\bibitem{PRD2019} M.K.-H. Kiessling, Phys. Rev. D {\bf 100}, 065012 (2019) (Erratum: Phys. Rev. D, {\bf 101}, 109901 (2020)).

\bibitem{PLA2013} F. Azzurli and K. Lechner, Phys. Lett. A {\bf 377}, 1025 (2013).

\bibitem{AP2014} F. Azzurli and K. Lechner, Ann. Phys. {\bf 349}, 1 (2014).







\bibitem{DasTQC} A. Das, {\it Lectures on Quantum Field Theory}, World Scientific, Singapore (2008).

\bibitem{Itzykson} C. Itzykson, J.B. Zuber, {\it Quantum Field Theory}, McGraw-Hill Inc, New York (1980).

\bibitem{Barcelos} J. Barcelos Neto, {\it Teoria de Campos e a Natureza: Parte Quântica}, Editora Livraria da F\'\i sica, S\~ao Paulo (2017).

\bibitem{GreinerQED}  W. Greiner, J. Reinhart, {\it Quantum Electrodynamics}, Springer, New York (2003).

\bibitem{Kaku} M. Kaku, {\it Quantum Field Theory: A Modern Introduction}, Oxford University Press, New York (1993).

\bibitem{Weinberg} S. Weinberg, {\it The Quantum Theory of Fields}, Vol. 1, Cambridge University Press, Melbourne (1995).

\bibitem{DasEletro} A. Das, {\it Lectures on Electromagnetism}, World Scientific, Singapore (2013).

\bibitem{Zangwill} A. Zangwill, {\it Modern Electrodynamics}, Cambridge University Press, Cambridge (2012).

\bibitem{Jackson} J.D. Jackson, {\it Classical Electrodynamics}, John Wiley \& Sons, New York (1999).

\bibitem{Frenkel} J. Frenkel, {\it Princ\'\i pios de Eletrodin\^amica Cl\'assica}, EdUSP, S\~ao Paulo (2005).

\bibitem{Felsager} B. Felsager, {\it Geometry, Particle, and Fields}, Springer, New York (1998).

\bibitem{BaroneHelayel} F.A. Barone, J.A. Helay\"el-Neto, Adv. Studies Theor. Phys. {\bf 2} (2008).

\bibitem{CamiloBaroneHelayel} F.A. Barone, G.T. Camilo, J.A. Helay\"el-Neto, Braz. J. Phys. {\bf 42}, 120 (2012).



\bibitem{qftdist} W. G{\"u}ttinger, Phys. Rev. {\bf 89}, 1004 (1953).

\bibitem{BJP2021} C.L.R. Braga and M. Sch\"onberg, Braz. J. Phys. {\bf 51}, 1276 (2021).

\bibitem{BJPamaku} M. Amaku, F.A.B. Coutinho, O.J.P. \'Eboli and E. Massad, Braz. J. Phys. {\bf 51}, 1324 (2021). (Erratum: Braz. J. Phys. {\bf 52}, 76 (2022)).




\bibitem{historydist} J. Liitzen, {\it The Prehistory of the Theory of Distributions}, Springer, New York (1982).

\bibitem{heavisidehist} B. L. Robertson, Trans. AIEE, {\bf 54}, 10 (1935).

\bibitem{deltazerotoinf} M. Amaku, F.A.B. Coutinho, O. \'Eboli and E. Massad, Rev. Bras. Ensino Fis., {\bf 43}, e20210132 (2021).

\bibitem{laurent} L. Schwartz, {\it Th\'eorie des distributions}, Hermann, Paris (1997).

\bibitem{SMA15} J. Alvarez, Sur. Math. Appl. {\bf 15}, 1 (2020).

\bibitem{colombeauLivro} J.F. Colombeau, {\it New Generalized Functions and Multiplication of Distributions}, Elsevier, Amsterdam (2000).

\bibitem{colombeau} J. F. Colombeau, [arXiv:0705.2396] (2007).





\bibitem{egorov} Y. V Egorov, Russ. Math. Surv., {\bf 45}, 1 (1990).

\bibitem{vickers} J.A. Vickers, J. Geom. Phys., {\bf 62}, 3 (2012).

\bibitem{steinbauer} R. Steinbauer and J. A. Vickers, Class. Quantum Grav., {\bf 23}, R91 (2006).

\bibitem{dannon} H. V. Dannon , Gauge Inst. Journal, {\bf 8}, 1 (2012).




\end{thebibliography}
\end{document}